\documentclass{aa}

\usepackage{txfonts}

\usepackage[english]{babel}
\usepackage{amsmath,amssymb}
\usepackage{graphicx, subfig}
\usepackage[normalem]{ulem}

\usepackage{verbatim}

\usepackage{multirow}
\usepackage{mathtools}
\usepackage{epstopdf}

\usepackage{url}
\usepackage{xcolor,color}
\usepackage{orcidlink}

\usepackage{natbib,twoopt}
\usepackage[hyphenbreaks]{breakurl}
\usepackage{booktabs}
\usepackage{placeins}

\bibpunct{(}{)}{;}{a}{}{,}             
\definecolor{cobalt}{rgb}{0.06, 0.2, 0.65}
\hypersetup{
  colorlinks,
  citecolor=cobalt,
  linkcolor=[rgb]{0.8, 0.2, 1.0},
  urlcolor=cobalt,
}
\makeatletter
  \newcommandtwoopt{\citeads}[3][][]{\href{http://adsabs.harvard.edu/abs/#3}%
    {\def\hyper@linkstart##1##2{}%
     \let\hyper@linkend\@empty\citealp[#1][#2]{#3}}}
  \newcommandtwoopt{\citepads}[3][][]{\href{http://adsabs.harvard.edu/abs/#3}%
    {\def\hyper@linkstart##1##2{}%
     \let\hyper@linkend\@empty\citep[#1][#2]{#3}}}
  \newcommandtwoopt{\citetads}[3][][]{\href{http://adsabs.harvard.edu/abs/#3}%
    {\def\hyper@linkstart##1##2{}%
     \let\hyper@linkend\@empty\citet[#1][#2]{#3}}}
  \newcommandtwoopt{\citeyearads}[3][][]%
    {\href{http://adsabs.harvard.edu/abs/#3}
    {\def\hyper@linkstart##1##2{}%
     \let\hyper@linkend\@empty\citeyear[#1][#2]{#3}}}
\makeatother

\def\apjl{ApJL}
\def\apjs{ApJS}
\def\aap{A\&A}

\def\be{\begin{equation}} 
\def\ee{\end{equation}} 
\def\ba{\begin{eqnarray}} 
\def\ea{\end{eqnarray}}

\def\msun{{\Msun}}

  \def\cm{${\rm cm}$}

\def\gsim{\lower.5ex\hbox{\gtsima}} 
\def\lsim{\lower.5ex\hbox{\ltsima}} \def\gtsima{$\; \buildrel > \over 
\sim \;$} \def\ltsima{$\; \buildrel < \over \sim \;$} \def\prosima{$\; 
\buildrel \propto \over \sim \;$} \def\gsim{\lower.5ex\hbox{\gtsima}} 
\def\lsim{\lower.5ex\hbox{\ltsima}} 
\def\simgt{\lower.5ex\hbox{\gtsima}} 
\def\simlt{\lower.5ex\hbox{\ltsima}} 
\def\simpr{\lower.5ex\hbox{\prosima}}   
  
 \def\gtsima{$\; \buildrel > \over \sim \;$} 
\def\ltsima{$\; \buildrel < \over \sim \;$} 
\def\gsim{\lower.5ex\hbox{\gtsima}} 
\def\lsim{\lower.5ex\hbox{\ltsima}} 
\def\simgt{\lower.5ex\hbox{\gtsima}} 
\def\simlt{\lower.5ex\hbox{\ltsima}} 
\def\simpr{\lower.5ex\hbox{\prosima}}

\def\msun{\,{\rm \Msun}}

\def\E3{{\cal E}_{\rm g}^{III}}

\def\Lsun{\rm L_\odot}

\def\r12{r_{1/2}} 
\def\x12{x_{1/2}} 

\def\cm2g{ {\rm cm^2} \, {\rm {g^{-1}}}}

\newcommand\code[1]{\textsc{\MakeLowercase{#1}}}
\newcommand{\quotes}[1]{``#1''}

\def\nh2{n_{\rm H2}}
\def\fh2{f_{\rm H2}}

\def\angstrom{\textrm{A\kern -1.3ex\raisebox{0.6ex}{$^\circ$}}}

\def\msun{{\rm M}_{\odot}}

\def\TdMWS20{${\langle T_{\rm d} \rangle}_{\rm M}^{\rm S20}$}
\def\TdLWS20{${\langle T_{\rm d} \rangle}_{\rm L}^{\rm S20}$}

\makeatletter
\def\@hex@@Hex#1%
 {\if a#1A\else \if b#1B\else \if c#1C\else \if d#1D\else
  \if e#1E\else \if f#1F\else #1\fi\fi\fi\fi\fi\fi \@hex@Hex}
\makeatother

\definecolor{apcolor}{HTML}{b3003b}
\definecolor{vdcolor}{HTML}{ff0f00}
\definecolor{afcolor}{HTML}{b3443c}
\definecolor{sgcolor}{HTML}{008000}
\definecolor{sccolor}{HTML}{ffbe33}
\definecolor{fvcolor}{HTML}{ce33ff}
\definecolor{pbcolor}{HTML}{333300}
\definecolor{fdcolor}{HTML}{ff6600}
\definecolor{ddcolor}{HTML}{26adf0}
\definecolor{todo}{HTML}{bc4be9}

\usepackage{siunitx}
\usepackage{cleveref}
\usepackage{xspace}

\crefformat{equation}{eq.~#2#1#3}
\crefformat{figure}{Fig.~#2#1#3}
\crefformat{section}{Sec.~#2#1#3}
\crefformat{subsection}{Sec.~#2#1#3}
\crefformat{table}{Tab.~#2#1#3}

\newcommand{\dd}{\mathop{\mathrm{d}\!}{}}
\newcommand{\deriv}[2]{\dfrac{\dd #1}{\dd #2}}

\newcommand{\ped}[1]{_{\rm #1}}
\newcommand{\lcdm}{$\Lambda$CDM\xspace}
\def\muv{{\rm M_{UV}}}
\def\luv{{\rm L_{UV}}}

\newcommand{\fpbh}{f\ped{PBH}}
\newcommand{\mpbh}{m\ped{PBH}}

\newcommand{\Mpbh}{M\ped{PBH}}

\DeclareMathOperator\erf{erf}

\graphicspath{{plots/}}

\linenumbers
\renewcommand\makeLineNumber{}

\begin{document}

\title{Can primordial black holes explain the overabundance of \\bright super-early galaxies?}
\titlerunning{Can PBHs explain the overabundance of bright super-early galaxies?}

\author{
Antonio Matteri$^{1}$ \orcidlink{0009-0007-9985-9112} \and
Andrea Pallottini$^{1,2}$ \orcidlink{0000-0002-7129-5761} \and
Andrea Ferrara$^{1}$ \orcidlink{0000-0002-9400-7312}
       }
\authorrunning{Matteri et al.}
\institute{
Scuola Normale Superiore, Piazza dei Cavalieri 7, 56126 Pisa, Italy
\and Dipartimento di Fisica ``Enrico Fermi'', Universit\'{a} di Pisa, Largo Bruno Pontecorvo 3, Pisa I-56127, Italy
}
\date{Received Jan 8, 2025; accepted XX XXX, XXXX}

\abstract
{
\textit{JWST} is detecting an excess of high-redshift ($z\gtrsim 10$), bright galaxies challenging most theoretical predictions.
To address this issue, we investigate the impact of Primordial Black Holes (PBHs) on the halo mass function and UV luminosity function (LF) of super-early galaxies.
We explore two key effects:
(i) the enhancement of massive halos abundance due to the compact nature and spatial distribution of PBHs, 
and (ii) the luminosity boost, characterized by the Eddington ratio $\lambda_E$, due to Active Galactic Nuclei (AGN) powered by matter accretion onto PBHs.
We build an effective model, calibrated using data at lower redshifts ($z\approx 4-9$), to derive the evolution of the LF including the additional PBH contribution.
Via Bayesian analysis, we find that:
(a) Although a small fraction ($\log \fpbh \approx -5.42$) of massive ($\log \Mpbh / \msun \approx 8.37$), non-emitting ($\lambda_E=0$) PBHs can explain the galaxy excess via the halo abundance enhancement, this solution is excluded by CMB $\mu$-distortion constraints on monochromatic PBHs.
(b) If PBHs power an AGN emitting at super-Eddington luminosity ($\lambda_E \approx 10$), the observed LF can be reproduced by a PBH population with characteristic mass $\log \Mpbh / \msun \approx 3.69$ constituting a tiny ($\log\fpbh \approx -8.16$) fraction of the cosmic dark matter content.
In the AGN scenario, about 75\% of the observed galaxies with $\muv=-21$ at $z=11$ should host a PBH-powered AGN and typically reside in low mass halos, $M_h = 10^{8-9} \msun$. These predictions can be tested with available and forthcoming \textit{JWST} spectroscopic data. We note that our analysis considers a lognormal PBH mass function and compares its parameters with monochromatic limits on PBH abundance. Further work is required to relax such limitations.
}


\maketitle

\section{Introduction} \label{sec:intro}

Since its launch, the \textit{James Webb Space Telescope} (JWST) led to the photometric discovery \citep{naidu:2022, castellano:2022, atek:2023, labbe:2023} and spectroscopical analysis \citep{arrabal:2023, bunker:2023, curtis:2023, robertson:2023, wang:2023, hsiao:2024,  zavala:2024} of galaxies up to about redshift $z=14$ \citep{carniani:2024}, when the Universe was $\approx 300\,\rm Myr$ old.

The abundant new data from \textit{JWST} are shedding light on the primordial Universe, allowing us to study the first galaxies in great detail. The analysis of the observations has shown the existence of an excess of high-redshift ($z\gtrsim 10$), bright galaxies challenging most theoretical predictions \citep{labbe:2023, robertson:2023, robertson:2024, casey:2024, harvey:2025, finkelstein:2024, fujimoto:2024}.

The inferred abundance of these systems is indeed about an order of magnitude higher than any prediction based on pre-\textit{JWST} data, such as hydrodynamical simulations (e.g. IllustrisTNG, \citealt{vogelsberger-nelson:2020}; FLARES, \citealt{vijayan:2021}; BlueTides, \citealt{wilkins:2017}), abundance matching models (e.g. UniverseMachine, \citealt{behroozi:2020}), and also extrapolations from lower redshift fits \citep[e.g.][]{bouwens:2022} of the UV luminosity function (LF).
Such tension constitutes an important challenge to our knowledge of the galaxy formation and evolution process in the early Universe, and several possible solutions have been proposed to solve this problem.

Astrophysical solutions include a stochastic star formation rate (SFR) in early galaxies that could boost the bright end of the LF to the observed level \citep{mason:2023, shen:2024}. However,
(i) the magnitude of the SFR flickering predicted from simulations is debated \citep{pallottini:2023, sun:2023},
(ii) the r.m.s. amplitude necessary to explain the overabundance problem is shown to be inconsistent with the observed mass-metallicity relation \citep{pallottini:2024}, and
(iii) spectral energy density analysis shows that the SFR stochasticity of early galaxies is relatively low \citep{ciesla:2024}.

Alternatively, the tension can be solved by scenarios in which either a negligible dust attenuation produced by radiation-driven outflows brightens the galaxies \citep[attenuation-free, ][]{ferrara:2023, ferrara:2024_outflow}, or the star formation efficiency is high due to inefficient supernova feedback in dense regions \citep[][]{dekel:2023}. These models and their implications are still under scrutiny.

Moreover, modifications of the \quotes{concordance} \lcdm model have also been explored. Indeed, if the number density of massive halos were higher than predicted by \lcdm, the tension could be released without postulating significant changes in the underlying astrophysics (e.g. star formation efficiency, dust attenuation, etc.). This can be achieved, for instance, by assuming different cosmological scenarios, such as an early dark energy contribution at $z\approx3500$ \citep{klypin:2021, shen:2024} that significantly changes the halo abundance at $z>10$.
In addition, the Halo Mass Function (HMF) can also be enhanced by (i) an effective modification of the transfer function \citep{padmanabhan:2023},
(ii) the presence of non-Gaussianities that make overdense regions more frequent \citep{biagetti:2023}, or
(iii) the contribution of primordial black holes (PBHs) to the halo formation \citep{liu:2022}. 

However, these proposals might cause other tensions: (i) an enhanced transfer function is argued to be excluded by low-$z$ HST observations \citep{sabti:2024}, (ii) the amplitude of non-Gaussian fluctuations is tightly constrained by CMB observations \citep{planckng:2020}, (iii) a monochromatic $\approx 10^{10}\,\msun$ PBH solution contributing to structure formation is excluded by Cosmic Microwave Background (CMB) $\mu$-distortions \citep{nakama:2018, gouttenoire:2024}.

In this work, we focus on the effects of PBHs on the observed galaxy LF relaxing the constraints imposed in \citet{liu:2022} by considering the possibility that PBHs have an extended mass function , as predicted by some inflationary models \citep[][]{garcia-bellido:2021, carr-clesse:2021}, and/or non-negligibly contribute to the UV luminosity of the host galaxy.
These two additions make the model more realistic, in principle allowing for less massive PBHs to have an impact on observations without violating current CMB constraints \citep[for the $\mu$-distortion limit in particular, see][]{wang:2025}\footnote{
    Throughout the paper we assume a flat Universe with \citet{planck:2018} parameters $\Omega_m = 0.30966$, $\Omega_{\Lambda}=0.68885$, $\Omega_{b}=0.04897$, $\sigma_8 = 0.8102$, and $h=0.6766$, where $\Omega_m$, $\Omega_{\Lambda}$, $\Omega_{b}$ are the matter, dark energy and baryon density ratios to the critical density, $h$ is the Hubble constant in units of $\SI{100}{\km \per \s \per \rm Mpc}$ and $\sigma_8$ is the fluctuations' r.m.s. amplitude parameter.
}.

\section{Methods} \label{sec:method}

We start by fixing the astrophysical model for star formation, emission, and attenuation by matching low ($z\approx4-9$) redshift data before considering the potential effects of PBHs.
To this aim, we compute the HMF for a given PBH mass distribution (\cref{sec:pbh-ps}),
fit the stellar emission of galaxies to the low-$z$ LF (\cref{sec:stellar-emission}),
add PBH emission contribution to the total UV luminosity (\cref{sec:pbh-emission}), compute the theoretical LF when PBHs are present (\cref{sec:pbh-distribution}), 
and then determining the main parameters within a Bayesian framework (\cref{sec:bayesian_determination}).

\subsection{PBH effects on structure formation}\label{sec:pbh-ps}

Following \citet{inman:2019}, the power spectrum of \lcdm can be modified through an additional term to account for the Poissonian shot noise \citep{carr:2018} produced by the discrete nature of PBHs
\begin{subequations}\label{eq:power-spectrum-pbh}
\begin{equation}
    P_{\rm CDM} = P_{\Lambda \rm CDM} + P_{\rm PBH}\,,
\end{equation}
with
\begin{equation}\label{eq:power-poisson}
    P_{\rm PBH} = \frac{f_{\rm PBH}^2}{n_{\rm PBH}}D^2\Theta(k_{crit}-k)\,,
\end{equation}
where $n_{\rm PBH}$ is the comoving number density of PBHs, $D=D(z)$ is the growth factor\footnote{
    Since only PBHs have a Poisson noise term, in the initial perturbations isocurvature modes should be accounted for. The growth factor from \citet{inman:2019} refers to these fluctuations instead of the usual adiabatic ones.}
from \citet{inman:2019}, and $f_{\rm PBH}$ can be written as
\be\label{eq:pbh-fraction}
    f_{\rm PBH} = \frac{n_{\rm PBH} m_{\rm PBH}}{\Omega_{dm}\rho_c}
\ee
for a monochromatic mass function.
$\Theta$ is the step function that suppresses the PBH contribution on scales smaller than the critical one \citep{liu:2022}
\be
k > k_{crit} = (2\pi^2 n_{\rm PBH}/f_{\rm PBH})^{1/3}\,.
\ee
Such scales are strongly affected by nonlinear evolution, due to the seed effect \citep{carr:2018}, and mode mixing between
isocurvature and adiabatic modes \citep{liu-zhang:2022}.
\end{subequations}
We normalize the power spectrum in \cref{eq:power-spectrum-pbh} by imposing the value for $\sigma(R=8\,h^{-1}\si{Mpc}) = \sigma_8$ from \citet{planck:2018}.

Following the excursion set theory for the halo collapse and using the approach from \citet{press:1974} accounting for the ellipsoidal collapse \citep{sheth:2002}, we compute the comoving number density of DM halos (Halo Mass Function, HMF) as a function of $z$ and of the halo mass ($M_h$).

Eqs. \ref{eq:power-spectrum-pbh} hold for PBHs with a monochromatic mass function, as considered by \citet{liu:2022}. In this work, we allow PBHs to have an extended mass function, adopting a lognormal distribution
\be\label{eq:lognormal-mf-pbh}
    \deriv{n(>M)}{\log M} \propto \frac{1}{\sqrt{2\pi}\sigma} \exp\left[-\frac12\left(\frac{\log M/M_{\rm PBH}}{\sigma}\right)^2\right]\,,
\ee
where $M_{\rm PBH}$ is the characteristic mass of PBHs, and $\sigma$ is the r.m.s. of the distribution; the constant of proportionality is fixed by normalizing the integral of the distribution to $f_{\rm PBH}$. The monochromatic case can be retrieved in the limit $\sigma \to 0$.

The lognormal mass function for PBHs is predicted by several inflationary models \citep[e.g.][]{dolgov:1993,green:2016,kannike:2017} and we choose it for its simple analytical form with only 3 parameters.
Other mass functions are produced when considering different models \citep[e.g.][]{garcia-bellido:2021},
eventually spanning a wider mass range than the lognormal and/or having several peaks.
Further investigation should be undertaken to test them within our framework.

\begin{figure}
    \centering
    \includegraphics[width=0.49\textwidth]{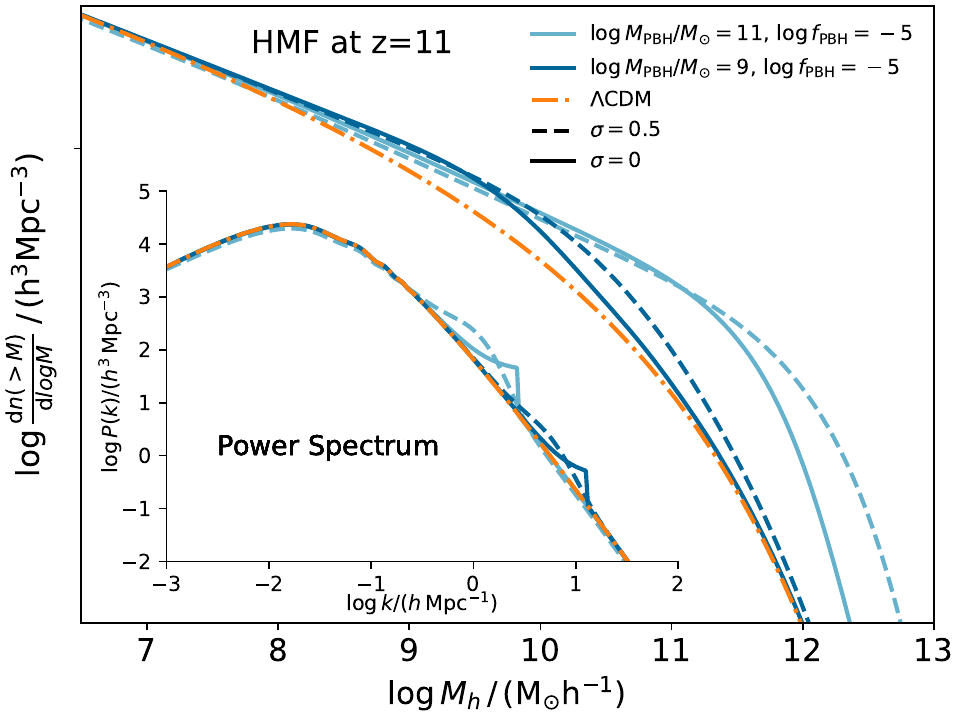}
    \caption{
    Overview of the modifications to the Halo Mass Function (HMF) due to the Primordial Black Holes (PBH) contribution to the power spectrum.
    The \citet{sheth:2002} HMF ($n$) at $z=11$ is plotted as a function of halo mass ($M_h$) for a pure \lcdm cosmology and for PBHs with lognormal mass functions,
    constituting a fraction $f_{\rm PBH}= 10^{-5}$ of the total dark matter density; each color corresponds to a different PBH mass $M_{\rm PBH}=10^9\,\msun,\, 10^{11}\,\msun$, whereas
    continuous (dashed) lines represent monochromatic (lognormal with $\sigma =0.5$) mass functions. 
    For reference, we show the corresponding power spectra at $z=0$ ($P_{\rm CDM}$, \cref{eq:power-spectrum-pbh}) as a function of the wavenumber ($k$) as an inset.
    \label{fig:hmf-lcdm-modication}
    }
\end{figure}

In \cref{fig:hmf-lcdm-modication} we show the HMF at $z=11$ for several PBH models to highlight the differences to the standard \lcdm structure formation. The corresponding power spectra are shown in the inset. Although in this case, PBHs constitute only a small fraction $\fpbh=10^{-5}$ of the dark matter density, they can significantly affect the power spectrum with a boost of $\approx 10\times$ around $k_{crit}$.
The width of the mass function changes the sharpness of the power spectrum contribution, i.e. the higher $\sigma$ the smoother it is around $k_{crit}$. An important feature of the power spectrum is the bump shift toward lower $k$ for higher $\sigma$ values. This is caused by
the extended tail of the lognormal distribution.
The general effect on the HMF is an enhancement in the halo abundance on a wide mass range peaked at a mass scale slightly larger than $\Mpbh$. More massive PBHs require a smaller $\fpbh$ to imprint the same relative increase in the halo abundance around $\Mpbh$.

The extended mass function changes the functional form of the power spectrum term in \cref{eq:power-poisson}, which must now be replaced with that in \cref{eq:power-lognormal}. In particular, the $\Theta$ function becomes smoother and has a width that depends on $\sigma$; the explicit derivation is given in \cref{app:lognormal-calculations}.

\subsection{Stellar emission}\label{sec:stellar-emission}

We build an effective model for the stellar emission from galaxies that matches pre-\textit{JWST} LF data and sets our baseline for super-early galaxy applications.
The LF ($\Phi_{\rm UV}$) as a function of $z$ is computed from the halo mass function $n(>M_h,z)$ through the chain rule 
\be\label{eq:luminosity-function}
    \Phi_{\rm UV} = \deriv{n}{\muv}(\muv, z) = \deriv{n(>M_h,z)}{M_h}\frac{1}{\deriv{\muv(M_h, z)}{M_h}}\,,
\ee
where $M_h$ is the mass of halos hosting galaxies of magnitude $\muv$. 
With this approach, we only need an emission model $\muv=\muv(M_h, z)$, i.e. the expected magnitude of a galaxy based on its halo mass.
Note that implicitly this neglects possible stochasticity contributions to the galaxy emission \citep[see e.g.][for more details on the impact]{mason:2023, pallottini:2023, gelli:2024, sun:2024}.

In this work, we choose a flexible functional form for the emission model
\be\label{eq:functional-form}
    \muv = p_1 + \left[p_2 \left(\frac{M_h}{10^{11} \msun}\right)^{p_3} + p_4\right] \log\left(\frac{M_h}{10^{11} \msun}\right) + p_5 z^{p_6}\,
\ee
where $p_i$ are parameters. The function combines a power-law with a logarithm to properly match the LF at fixed redshift with an additional term
for the explicit dependence on $z$. Despite this particular choice lacking a clear physical motivation, we find it gives good results over the whole redshift range
we are considering.
We fit these parameters to the LF data at $z = 4,\,5,\,6,\,7,\,8,\,\mathrm{and}\,9$ from \citet{bouwens:2021}, by assuming a purely \lcdm model ($\fpbh=0$ in \cref{eq:power-spectrum-pbh}) when computing the LF (\cref{eq:luminosity-function}). Errors on LF data are treated as Gaussian,
which is a rough approximation, but allows us to account for uncertainties of the fit in a simple way. The result is given by
\be\label{eq:fit-results-muv-vs-mh}
    p_{\rm fit} =
    \begin{pmatrix}
        p_1 \\
        p_2 \\
        p_3 \\
        p_4 \\
        p_5 \\
        p_6
    \end{pmatrix}_{fit}
    =
    \begin{pmatrix}
        -13.58\\
        -8.80\\
        0.0751\\
        5.68\\
        -3.44\\
        0.299
    \end{pmatrix}
    \pm
    \begin{pmatrix}
        5.91 \\
        7.67 \\
        0.0683 \\
        7.67 \\
        5.35 \\
        0.306
    \end{pmatrix}\,
\ee
where the error of the fit is relatively large due to the high correlation between the parameters
retrieved from the 6 parameters least square error procedure.

\begin{figure}
    \centering
    \includegraphics[width=0.49\textwidth]{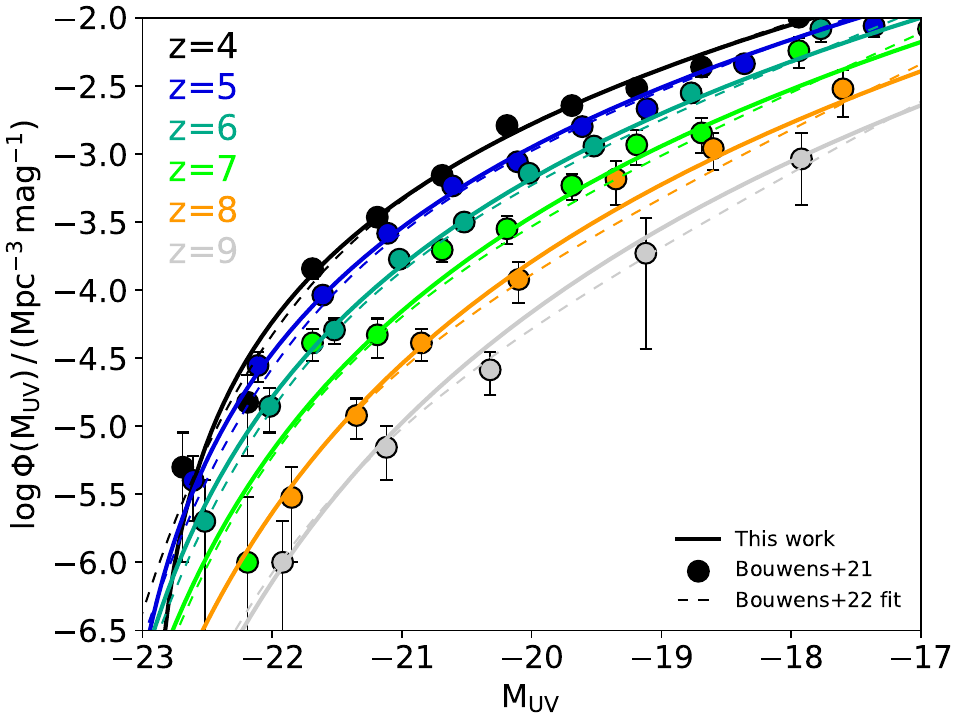}
    \caption{
        UV luminosity function of galaxies at different redshifts compared to data
        \citep{bouwens:2021, bouwens:2022}.
        The LF from our model (solid line, \cref{eq:functional-form}) is fitted to the points at the reported redshifts (parameters in \cref{eq:fit-results-muv-vs-mh}).
        \label{fig:lum-fun-fit}
    }
\end{figure}

In \cref{fig:lum-fun-fit} we report the modeled LF at different $z$ along with the fitting data \citep{bouwens:2021} and the Schechter function fit\footnote{We underline that the fit from \citet{bouwens:2022} is performed directly on the LF, while we perform a fit on the $\muv-M_h$ relation.} from \citet{bouwens:2022}.
The fitting function (\cref{eq:functional-form}) appears to provide a very good approximation to the data over the whole redshift range of interest. The plot has been produced using the best-fit values
from \cref{eq:fit-results-muv-vs-mh} and neglecting their uncertainties. In general, this is a bad practice since it does not provide information about the robustness of the results. However, in this context, we are not interested in the precise values of the parameters, but only in the model agreement with the data points. In other words, it is not a problem if the solution we find does not have tightly constrained parameters. Therefore, we proceed to neglect the uncertainties in 
\cref{eq:fit-results-muv-vs-mh}
and consider only the best-fit values in the following treatment.

\begin{figure}
    \centering
    \includegraphics[width=0.49\textwidth]{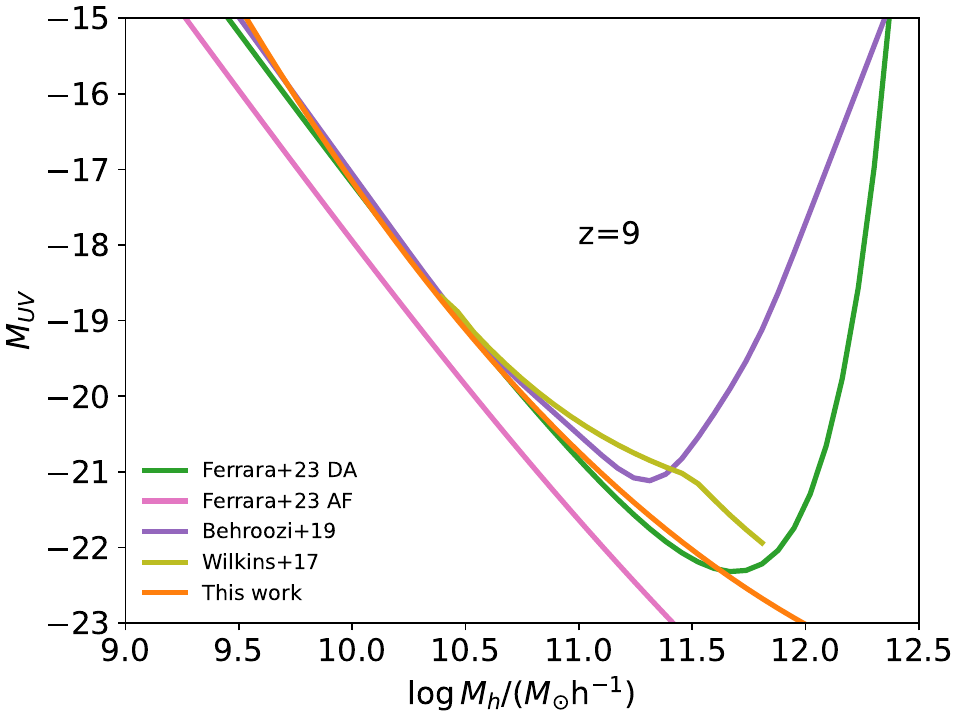}
    \caption{
        UV magnitude ($\muv$) as a function of $M_h$ at $z=9$ within different models.
        The relation in this work is obtained via a fit of the
        luminosity function (see \cref{eq:functional-form} for the functional form and \cref{eq:fit-results-muv-vs-mh}
        for the parameters).
        For comparison, we report $\muv(M_h)$ from the semi-analytical model of \citet[][in the dust-attenuated, DA, and attenuation-free, AF, cases]{ferrara:2023}, from the abundance-matching based procedure of \citet{behroozi:2019} and results from the BlueTides hydrodynamical simulations \citep{wilkins:2017}.
        \label{fig:muv-vs-mh}
    }
\end{figure}

We plot the $\muv-M_h$ relation at $z=9$ (\cref{fig:muv-vs-mh}) to compare our fiducial model with previous works.
The models from \citet{ferrara:2023} are built through a semi-analytical model obtained fitting the LF at $z=7$ and considering the dust attenuation \citep{inami:2022} from the REBELS survey \citep{bouwens:2022_rebels}; in addition, the attenuation-free case is reported labeled by AF.
\citet{behroozi:2019} model uses the stellar-halo mass relation, derived via abundance matching, combined with their eq. D5 to link the stellar mass to the UV magnitude, including dust attenuation.
We also report the results of the BlueTides \citep{wilkins:2017,feng:2015,feng:2016} hydrodynamical N-body simulation focused on the properties of galaxies with stellar masses $M_{\ast}\ge 10^8\,\msun$ at $z\ge8$. To compute the observed UV luminosity from simulated data they assume the intrinsic UV luminosity of a galaxy to be $\propto$ SFR, and its dust attenuation to scale linearly with metallicity.

Except for the attenuation-free case from \citet{ferrara:2023}, the $\muv-M_h$ relation from our model is very close to the ones from literature up to $M_h\approx 10^{11}\,\msun$.
Above this threshold models behave differently, and in our model the luminosity of a galaxy monotonically increases with halo mass. This indicates that we may be missing heavily dust-obscured galaxies (expected in dust-attenuated models, see \citealt[][]{ferrara:2023} [DA], and \citealt{behroozi:2019}) and eventually also Active Galactic Nucleus (AGN) feedback (considered in \citealt{wilkins:2017}). These effects, however, become important only for $M_{\rm UV} \lesssim -22$, and hence do not dramatically impact our conclusions.

We underline that we are missing detailed information on the galaxy obscuration since it is included implicitly in the fit done with \cref{eq:functional-form} \citep[cfr. with][]{ferrara:2023}. From \cref{fig:muv-vs-mh}, we can see that dust extinction affects only galaxies contained in massive ($\gtrsim 10^{12}\,\msun$), thus rare, halos. We treat the AGN dust extinction in the following section.

\subsection{PBH emission}\label{sec:pbh-emission}

Accreting PBHs can emit light as AGN, thus increasing the brightness of their host galaxies.
We quantify PBH emission via the Eddington ratio $\lambda_E$, defined as the ratio between the AGN bolometric luminosity, $L_{\rm bol}$, and its
Eddington luminosity:
\be\label{eq:eddington-luminosity}
    L_{E} = \frac{4\pi G m_p c}{\sigma_T}m_{\rm PBH} = 3.28\times10^4 \,\frac{m_{\rm PBH}}{\msun}\,\Lsun\,,
\ee
where $G$ is Newton's constant, $m_p$ the proton mass, $c$ the speed of light, and $\sigma_T$ the cross section for Thomson scattering.
From the bolometric luminosity ($L_{\rm bol} = \lambda_E L_{E}$), we retrieve the UV luminosity through the conversion factors from \citet{shen:2020}
\begin{subequations}
\be
    {\rm L_{UV, \,PBH}} = \frac{L_{\rm bol}} {c_1 \left(\frac{L_{\rm bol}}{ 10^{10}\,\Lsun}\right)^{k_1} + c_2 \left(\frac{L_{\rm bol}}{10^{10}\,\Lsun}\right)^{k_2}}
\ee
where:
\be
    \begin{pmatrix}
        c_1 \\
        k_1 \\
        c_2 \\
        k_2
    \end{pmatrix}
    =
    \begin{pmatrix}
        1.862\\
        -0.361\\
        4.870\\
        -0.0063
    \end{pmatrix}
\ee
and sum it to the stellar contribution to obtain the total emission from the galaxy:
\be
    {\rm L_{UV, \,tot}} = \luv(M_h, z) + {\rm L_{UV, \,PBH}}\, .
\ee
\end{subequations}
Note that we do not account explicitly for dust attenuation in the AGN term, which reduces the observed luminosity, rather
it is considered implicitly in the bolometric correction factors.
The total UV magnitude is used to compute the LF including PBH emission.
For simplicity, we assume that $\lambda_E$ is the same for each PBH, although more realistically we do expect this parameter to have a wide distribution \citep{Bhatt24}. The case in which PBHs do not emit can be retrieved by imposing $\lambda_E=0$.

We warn that we use the same PBH mass function to compute the modified power spectrum and the PBH emission term even though they should correspond to the one at matter-radiation equality and that at the observed $z$, respectively.
Stated differently, we neglect the accretion of PBHs from their formation epoch up to the observed $z$ \citep{jangra:2024,nayak:2012}. This implies that the required $\lambda_E$ values to fit the observed LF must be seen as upper limits. Assuming
a specific model for accretion \citep[e.g.][]{hasinger:2020} we can compute the PBH mass growth. 
Supposing that each of them
grows by a factor $K$ before $z=11$,
we retrieve the same results as considering a model in which PBH do not grow and
setting $\lambda_E^\prime = K \lambda_E$ (further discussion in \cref{sec:discussion}).

We underline that here we are not accounting for the AGN duty cycle. Values $<1$ for this quantity would reduce the number
of active AGN, thus requiring a higher number of PBHs (i.e. $\fpbh$) to retain the same results.

\subsection{PBH spatial distribution}\label{sec:pbh-distribution}
To properly estimate the contribution to the LF, we must populate galaxies with PBHs using some educated guesses.

First, we assume that all PBHs reside in galactic halos rather than in the intergalactic medium. This choice maximizes the impact of PBH emission at fixed $\fpbh$ and it is motivated by the notion that PBHs may foster the growth of baryonic structures around them. This point has been made by some works, such as e.g. \citet[][who conclude that isolated PBH are an artifact preferentially appearing for large $\fpbh$ values]{inman:2019}, while other ones consider the presence of some intergalactic PBHs \citep{manzoni:2024, jangra:2024}.

Next, we distribute PBHs in halos with a mass $M_h > M_{min}=10^{7.5} \,\msun$. This choice is motivated by the fact that it should be unlikely to find PBHs in small halos and that galaxies hosted by halos with $M_h < M_{min}$ are fainter than the lowest LF constraints (according to our astrophysical model, \cref{sec:stellar-emission}). With larger $M_{min}$ values instead the LF at $z=11$ cannot be reproduced properly varying the model parameters, since the predicted LF has a different behavior in the faint end than the data. Our recipe is admittedly rough and should be refined based on more stringent physical arguments in the future. This could be done along the lines initially explored in \citet{inman:2019} and \citet{liu-zhang:2022}.

In practice, we implement the previous assumption by imposing that halos with $M_h > M_{min}$ contain a number of PBHs drawn from a Poisson distribution with mean $x$, independent of $M_h$. Each PBH is assigned a mass by extracting from the mass function. The value of $x$ can be determined
by matching the total cosmic number density of PBHs,
\be
    x = \frac{f_{\rm PBH} \Omega_{dm} \rho_c }{n(>M_{min})\langle M_{\rm PBH}\rangle}
\ee
where $\langle M_{\rm PBH}\rangle$ is the mass function-averaged PBH mass, and $n(>M_{min})$ is the number density of halos with $M_h >M_{min}$.
Therefore, we split the halos with a certain mass in groups based on the number of PBHs they host, we compute the PBH UV luminosity distribution
within each group and add it to the stellar one.
In the end we bin the galaxies based on their UV luminosity to retrieve the LF.
With this approach, halos of the same mass may host objects with different $\muv$ due to the different contributions of PBHs, that
may vary in number (following a Poissonian distribution) and in mass (according to their mass function).

\subsection{Model parameters and their determination}\label{sec:bayesian_determination}

In summary, the model has 4 parameters (3 for the mass function and $\lambda_E$), which we constrain with the observed LF considering both the effects due to PBHs: the power spectrum enhancement and the additional UV emission.
We expect the model to present degeneracies since massive PBHs with low $\lambda_E$ emit exactly like smaller PBHs with a higher $\lambda_E$.
To avoid problems connected to the presence of such degeneracies, we work in a Bayesian framework imposing flat priors on the mass function parameters within the intervals:
\be\label{eq:priors}
\begin{aligned}
    -20  &\le& \log \fpbh \quad \; &\le& -2 \\
    1    &\le& \log \Mpbh / \msun &\le& 11 \\
    0.01 &\le& \sigma  \hspace{0.9cm} &\le& 1.5
\end{aligned}
\ee
and considering different values for $\lambda_E$ ($0$, $0.01$, $0.1$, $1$, and $10$). To obtain the posterior distribution, we run Monte Carlo Markov Chains (MCMCs) considering LF data at $z=11$ from \citet{mcleod:2024} and \citet{donnan:2024} as constraints.

\FloatBarrier
\section{Results} \label{sec:results}
\begin{figure*}[ht]
    \centering
    \includegraphics[width=0.99\textwidth]{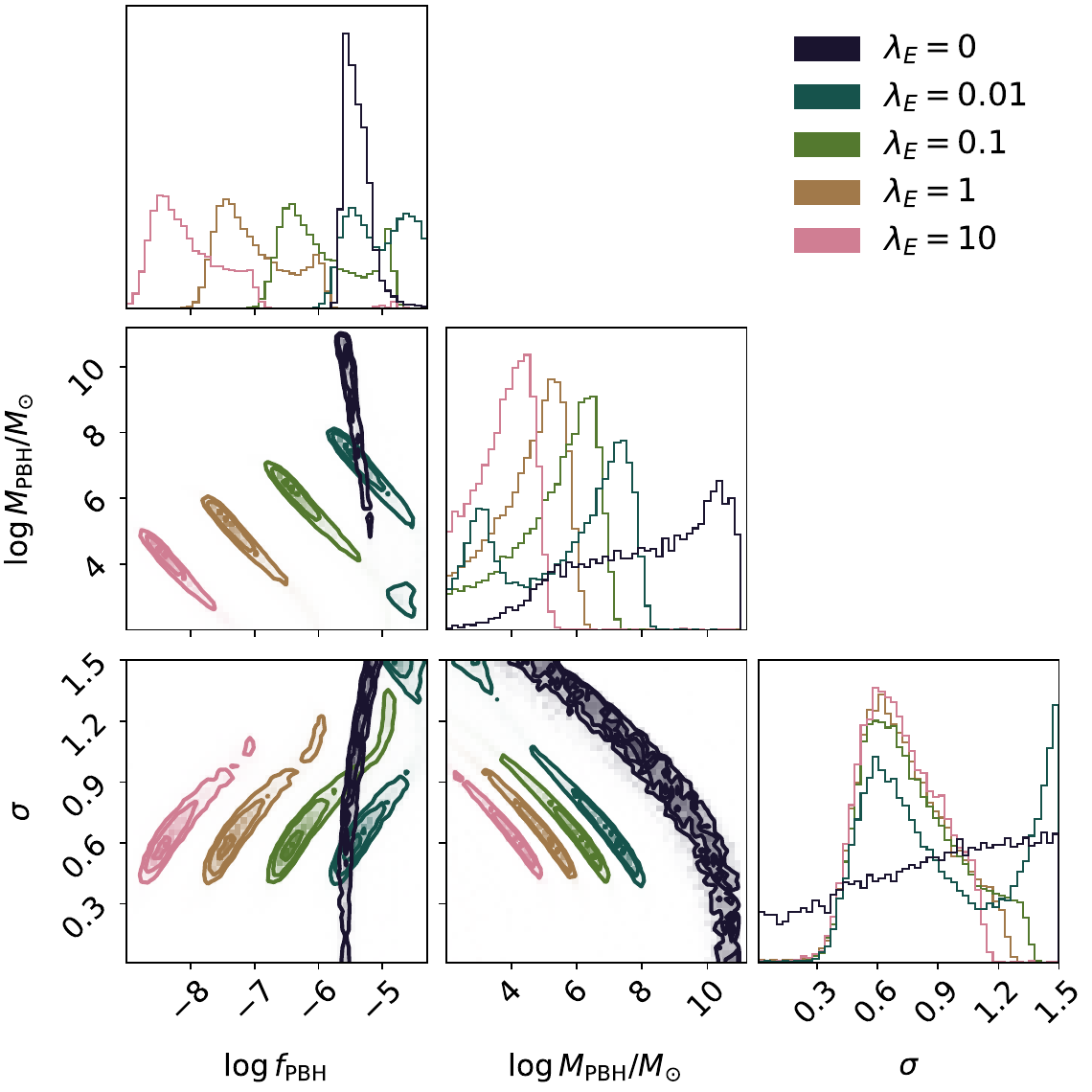}
    \caption{
    Corner plot of the posterior for the PBH models (\cref{sec:method}).
    The parameters describing the mass function of the PBHs (\cref{eq:lognormal-mf-pbh}) are obtained via the MCMC using constraints at $z=11$ and assuming PBHs are distributed independently of halo mass.
    Each color corresponds to a value for $\lambda_E$, as reported in the legend. An interesting feature to note is that, as long as $\lambda_E>0.01$, changing $\lambda_E$ shifts the posterior distribution, as indicated in \cref{eq:peaks-relation}.
    Tabulated results are reported in \cref{tab:emitting-pbh}.
    \label{fig:corner-plot}
    }
\end{figure*}

\begin{table}[t]
    \centering
    \begin{tabular}{c | c c c}
        \toprule
        $\lambda_E$ & $\log \fpbh$ & $\log \Mpbh / \msun$ & $\sigma$\\
        \midrule
        10   & $-8.16^{+0.73}_{-0.39}$ & $3.69^{+0.84}_{-1.43}$ & $0.70^{+0.23}_{-0.18}$ \\
        1    & $-7.12^{+0.87}_{-0.42}$ & $4.60^{+0.91}_{-1.71}$ & $0.71^{+0.27}_{-0.18}$ \\
        0.1  & $-6.06^{+0.99}_{-0.46}$ & $5.49^{+0.99}_{-2.08}$ & $0.73^{+0.31}_{-0.20}$ \\
        0.01 & $-4.90^{+0.55}_{-0.62}$ & $5.98^{+1.44}_{-3.01}$ & $0.83^{+0.57}_{-0.27}$ \\
        0    & $-5.42^{+0.22}_{-0.16}$ & $8.37^{+1.92}_{-2.68}$ & $0.93^{+0.39}_{-0.51}$ \\
        \bottomrule
    \end{tabular}
    \caption{
        Tabulated MCMC results for corner plot reported in \cref{fig:corner-plot}.
        For each parameter, the fiducial value corresponds to the median, the uncertainties are built using the 16 and 84 percentiles.
        \label{tab:emitting-pbh}
    }
\end{table}

The posterior distributions resulting from our Bayesian analysis are reported in \cref{fig:corner-plot} in the form of a corner plot and in \cref{tab:emitting-pbh} using the 16, 50, and 84 percentiles. 
Values of the parameters corresponding to different cases of $\lambda_E$ (10, 1, 0.1, 0.01, and 0) are reported separately
to easily compare the results. An alternative parametrization of the results, that may be useful when comparing to other works, is reported in \cref{app:second-corner}.

In the $\lambda_E=0$ case, very massive PBHs are required to match the LF since the power spectrum enhancement affects the HMF on a mass scale $\approx M_{\rm PBH}$ and we need to boost the abundance of galaxies that have $M_h\gtrsim10^{10}\,\msun$.
From the corner plot, we also see that broader PBH mass functions (larger $\sigma$) require a smaller $\Mpbh$ due to the skewed shape of the lognormal distribution.

For cases with $\lambda_E \gtrsim 0.01$, the shapes of the posteriors are very similar to each other, but the location appears shifted in $\Mpbh$ and $\fpbh$. We find that the peak position satisfies the \quotes{degeneracy conditions}
\begin{subequations}\label{eq:peaks-relation}
\be\label{eq:degeneracy-one}
    \Mpbh\lambda_E\approx10^{4.5}\,\msun
\ee
and
\be\label{eq:degeneracy-two}
    \Mpbh / \fpbh \approx 10^{11.7}\, \msun\,,
\ee
\end{subequations}
up to statistical uncertainties.
Compared to the non-emitting case, the required PBH masses are lower, and the posterior on $\sigma$ are no longer flat but rather peaked around $0.7$.

The different behaviors and shapes of the non-emitting and emitting solutions can be understood as follows.
When $\lambda_E=0$, by definition, only the power spectrum enhancement can induce changes in the LF.
For solutions with $\lambda_E>0.01$, the DM power spectrum is instead virtually the same as the \lcdm one, because of the small PBH mass; thus, the PBH-powered AGN luminosity contribution dominates the LF enhancement.

The relations in \cref{eq:peaks-relation} state that the number density of PBHs is fixed (see \cref{eq:pbh-fraction}) and that their luminosity distribution (see \cref{sec:pbh-emission}) is also the same. This holds until the power spectrum contribution breaks this degeneracy. 

The $\lambda_E=0.01$ case presents a second solution with a very broad mass function ($\sigma\approx1.5$) and lower typical mass ($\Mpbh\approx 10^3 \,\msun$), absent in other cases. This solution lies on the border of our prior distribution, meaning that we can get only partial information about such wide mass functions within the limitations of this work.
For such a case, the combination of the retrieved parameters takes advantage of both the power spectrum enhancement (in the faint end) and of the PBH emission (in the bright end) to recover the detected LF.

\begin{figure}[h!]
    \centering
    \includegraphics[width=0.49\textwidth]{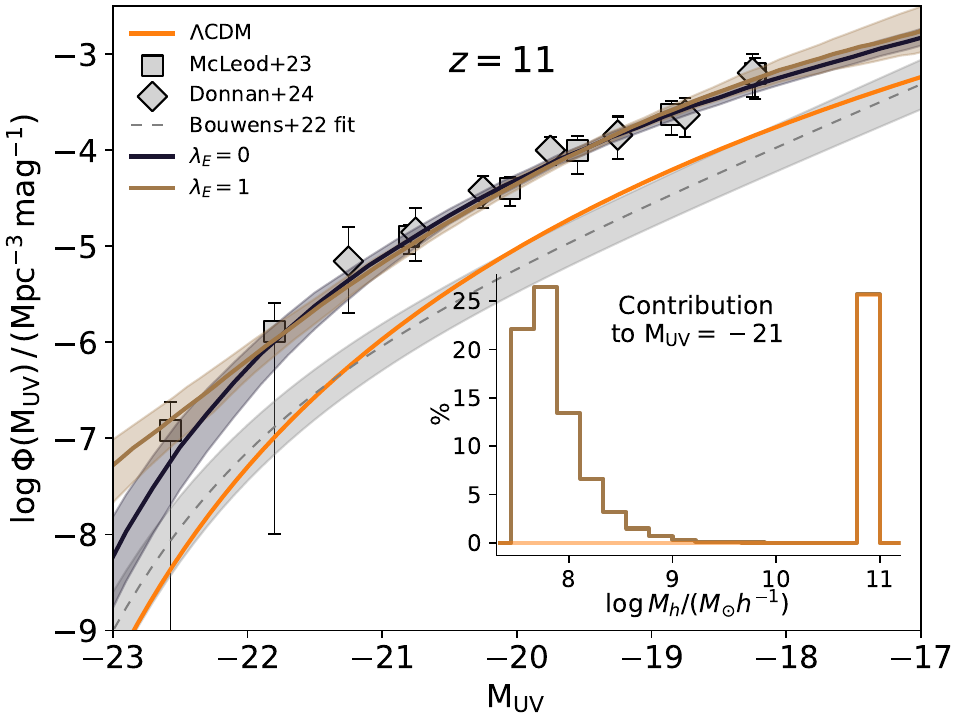}
    \caption{
    Predicted UV luminosity function ($\Phi_{\rm UV}$) at $z=11$.
    Our standard \lcdm model without PBHs (\cref{sec:stellar-emission}; solid orange curve) is compared with models including PBHs with $\lambda_E=0$ (solid black) and $\lambda_E=1$ (solid brown); shaded regions denote model uncertainties (16 and 84 percentiles at fixed $\muv$).
    Data from \citet{mcleod:2024} and \citet{donnan:2024} at $z=11$, and the fit from \citet{bouwens:2022} to low-$z$ data as a reference for our fiducial model are also shown.
    The inset shows the percentage of $\muv=-21$ galaxies that are hosted by halos of mass $M_h$ for the $\lambda_E=1$ case (brown line);
    the orange line instead is computed within the \lcdm model without changing the normalization.
    Although LF curves for both values of $\lambda_E$ agree with data, they differ in the bright end of the LF.
    \label{fig:lum-fun}
    }
\end{figure}

To visualize how the LF is affected by PBHs, in \cref{fig:lum-fun} we compare the LF predicted by our standard \lcdm model without PBHs (Sec. \ref{sec:stellar-emission}, with that including PBHs with $\lambda_E=0$ and $\lambda_E=1$. 
A striking difference between the two PBH models is evident at the LF bright end. The non-emitting case ($\lambda_E=0$) presents a rapidly decreasing LF, which is similar to the exponential tail of a Schechter function. On the other hand, the case in which PBHs are emitting at the Eddington rate ($\lambda_E=1$) predicts a much flatter slope at bright magnitudes, resembling a double power-law trend.
This is particularly interesting as there has been some debate in the literature concerning the appropriate functional form of the bright-end of the LF at $z\gtrsim7$ \citep{bowler:2014, donnan:2023}. In our model, the actual shape of the LF is determined by the accretion efficiency onto PBHs.

The inset in \cref{fig:lum-fun} shows the fraction of galaxies with $\muv=-21$ that are hosted by halos with mass $M_h$. 
In the standard \lcdm model (and also in the $\lambda_E=0$ case, not shown) the contribution comes from a single bin around $10^{11}\,\msun$.
For $\lambda_E=1$, instead, the distribution features a double-peak structure, with a high-mass (low-mass) peak produced by stellar (PBH-powered AGN) emission.
As there is no significant difference in the rightmost part of the histograms, the bright-end LF enhancement is therefore mainly driven by galaxies in $10^{8-9} \,\msun$ halos. 

\section{Discussion} \label{sec:discussion}

\begin{figure}[ht]
    \centering
    \includegraphics[width=0.49\textwidth]{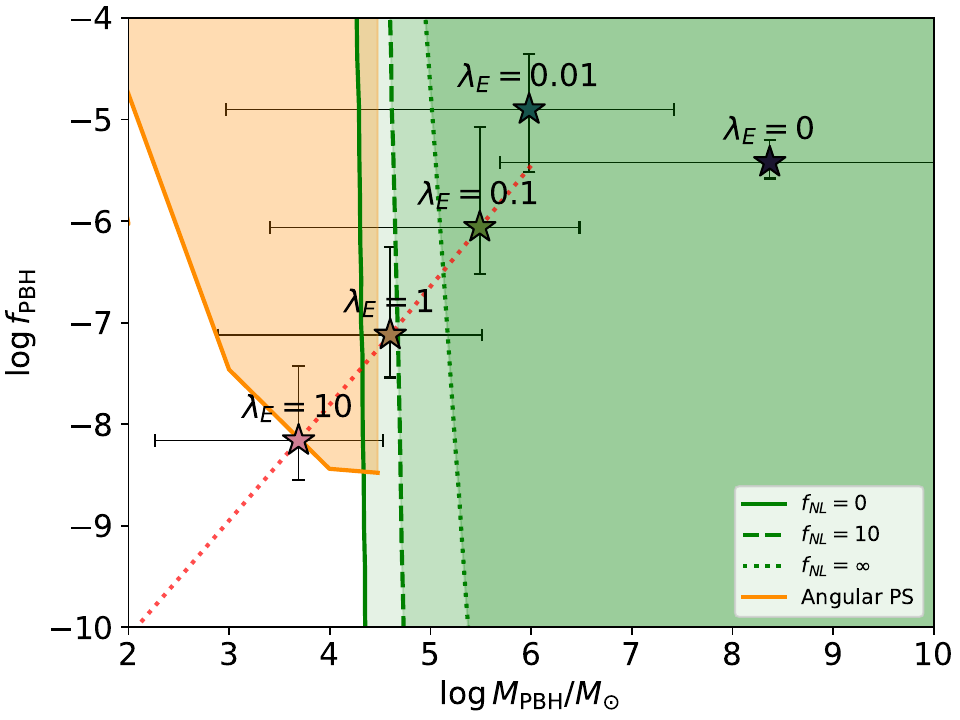}
    \caption{
        PBH parameter exclusion plot in the $\fpbh-\mpbh$ plane.
        We plot the constraint on monochromatic PBHs abundance from the CMB $\mu$-distortion absence (\citealt{nakama:2018},
        green lines) within different non-Gaussianity scenarios ($f_{\rm NL}=0,10,\infty$ represented by the solid, dashed,
        and dotted lines respectively) and the one from the observed CMB angular power spectrum (\citealt{serpico:2020}, orange line).
        Shaded regions are excluded and their colors denote the violated limit (same color scheme of the lines).
        Stars and error bars give the results and uncertainty we find for different values of $\lambda_E$ (\cref{tab:emitting-pbh}).
        The red dotted line traces the expected trend varying $\lambda_E$ in the case only
        AGN contribution is present.
    \label{fig:exclusion-plot}
    }
\end{figure}

PBHs have attracted considerable attention as possible dark matter candidates. These studies are limited in their predictions by the persisting ignorance of their typical mass, initial mass distribution, and abundance \citep{carr-kohri:2021}.

In \cref{fig:exclusion-plot} we report the currently available limits on monochromatic PBHs in the mass range relevant to the present study. Shaded regions correspond to combinations of parameters in tension with observations and stars represent our results. For simplicity, constraints assume a monochromatic mass function, while results are reported as the peak of the posterior distribution we find. Lognormal mass functions usually have tighter limits due to the long tail of the distribution \citep{bellomo:2018}.

Between $10^2\,\msun$ and $10^{10}\,\msun$ the two main constraints come from CMB observations. The first is from the undetected $\mu$-distortion \citep{nakama:2018}; the second is obtained from the angular power spectrum \citep{serpico:2020}.
The $\mu$-distortion is a deviation in the CMB temperature distribution that may be caused by local heating due to PBHs when the Universe age was between $7\times10^6\,\si{s}$ and $3\times10^{9}\,\si{s}$.
In \cref{fig:exclusion-plot}, the standard \lcdm limit is reported as $f_{\rm NL}=0$ (Gaussian fluctuations), while non-Gaussianity models are presented due to the loosening of constraints with higher $f_{\rm NL}$ (we plot the cases $f_{\rm NL} = 10,\,\infty$).
According to \citet{planckng:2020}, values of $f_{\rm NL}>10$ are incompatible with observations.
The CMB angular power spectrum limit is derived by computing the effect of accreting PBHs on the baryon angular distribution at the last scattering surface. Massive and abundant PBHs would indeed produce detectable patterns in the baryon density field.

The results we obtain in the case $\lambda_E=0$ are compatible with the ones from \citet{liu:2022}, who find that $\fpbh\mpbh\gtrsim 200\,\msun$, $\mpbh \gtrsim 5\times10^8\,\msun$ and $\sigma=0$ (they consider only monochromatic PBHs)
are required to explain \textit{JWST} observations without assuming a high star formation efficiency.
However, our $\lambda_E=0$ solution leads to two problems: i) at least a fraction of such massive black holes would probably accrete gas and emit with a detectable $\lambda_E>0$ and ii) the required abundance would be higher than the $\mu$-distortion limits for their mass even when using an extended mass function for PBHs.
Allowing for PBH emission ($\lambda_E > 0$) ameliorates both issues as it automatically addresses the former and reconciles the results with the $\mu$-distortion data constraints.

When considering high $\lambda_E$ values, the mass function resulting from our study shifts toward lower mass scales, evading the $\mu$-distortion constraint.
However, the angular power spectrum limit becomes dominant when $\Mpbh\lesssim 10^4\,\msun$ (see \cref{fig:exclusion-plot})
and values $\lambda_E\gtrsim10$ provide consistent solutions to the problem.

The second solution found in the $\lambda_E=0.01$ case is in tension with the CMB angular power spectrum
constraint even though much smaller PBH masses are required than in the other cases.
When comparing this solution with monochromatic limits, we must pay attention to the fact
that the mass function is very wide: PBHs with mass $\approx 10^6\,\msun$
are $2\sigma$ outliers and their sole abundance is 
incompatible with $\mu$-distortion limits, assuming that the constraint is
not weakened by abundant low-mass PBHs.

Exploring diverse mass functions and considering even broader
ones \citep[e.g.][]{garcia-bellido:2021} may provide useful insight on this topic and possibly find solutions compatible with CMB limits. Since results obtained by assuming extended mass functions are not reliably comparable with monochromatic constraints, a model-dependent analysis is required to solidly rule out the corresponding solutions. For example, \citet{wang:2025} report looser constraints from the $\mu$-distortion with
a wide enough ($\sigma\gtrsim0.3$) lognormal mass function when specific non-Gaussian primordial fluctuations models are assumed.

These considerations imply that the emission from PBHs acting as AGN is required for them to both explain the LF enhancement and comply with CMB observations.
Moreover, our model entails that in 75\% of the observed $\muv=-21$ sources at $z\approx 11$ the emission should be dominated by an AGN (see inset in \cref{fig:lum-fun}), making it testable against spectroscopic observations. So far, spectroscopy on $z\gtrsim10$ objects has revealed both AGN (e.g. UHZ1, \citealt{natarajan:2024}; GHZ2, \citealt{castellano:2024}) and star-forming galaxies (e.g. JADES-GS-z14-0, \citealt{carniani:2024}), but more data are required to provide a statistical sample to test our prediction.

We recall that we have ignored the accretion of PBHs from their formation up to redshift $z\approx 11$. To properly address this issue, further modeling of the PBH environment \citep{nayak:2012,jangra:2024} is required.
Naively, though, we note that if PBHs increase their mass of a factor $100$ before $z=11$ (as permitted by an efficient Bondi-Hoyle-Lyttleton accretion, see \citealt{jangra:2024}), they would explain \textit{JWST} observation with $\lambda_E$ being 100 times smaller then the ones found here (see \cref{fig:corner-plot}).

We also assumed that all PBH-powered AGN predicted by the model shine continuously and simultaneously, i.e. we have not attempted to model the likely possibility that their duty cycle is $< 1$. In that case, a higher $\fpbh$ would be required to match the LF since a fraction of the PBHs would be quiescent.
Unfortunately, the duty cycle would be degenerate with $\fpbh$. Detailed numerical simulations could be useful to break such a degeneracy. 

\section{Summary} \label{sec:conclusions}

We have investigated whether the inclusion of PBHs in the standard \lcdm model can alleviate the problems raised by the observed excess of bright, super-early ($z>10$) galaxies. 
To this aim we have computed the PBH contribution to the halo mass function (\cref{sec:pbh-ps}) and the galaxy LF (\cref{sec:pbh-emission}) for a lognormal initial PBH mass function (with a peak at $\Mpbh$ and amplitude $\sigma$, \cref{eq:lognormal-mf-pbh}), further assuming that they 
constitute a fraction $\fpbh$ of the dark matter and power an AGN with an Eddington ratio $\lambda_E$. Using a Bayesian analysis (\cref{sec:bayesian_determination}), we find that:
\begin{itemize}

\item[{\color{red} $\blacksquare$}] Although a small fraction ($\log \fpbh \approx -5.42$
) of massive ($\log \Mpbh / \msun \approx 8.37$ with $\sigma \approx 0.93$), 
non-emitting ($\lambda_E=0$) PBHs can explain the galaxy excess (\cref{fig:corner-plot}), this solution is in contrast with CMB $\mu$-distortion constraints on monochromatic PBHs (\cref{fig:exclusion-plot}) and therefore must be discarded.

\item[{\color{red} $\blacksquare$}] If PBHs power an AGN emitting at super-Eddington luminosities ($\lambda_E \approx 10$), the observed LF can be reproduced by a PBH population with characteristic mass $\log \Mpbh / \msun \approx 3.69$ constituting a tiny ($\log \fpbh \approx -8.16$) fraction of the cosmic dark matter content.

\item[{\color{red} $\blacksquare$}] As $\lambda_E$ and $\Mpbh$ are degenerate, the LF remains unchanged if the degeneracy condition $\Mpbh \lambda_E = \rm 10^{4.5}\ \msun$ is satisfied. A similar degeneracy exists between $\lambda_E$ and $\fpbh$ (\cref{eq:peaks-relation}).

\item[{\color{red} $\blacksquare$}] Although the LF can be reproduced by any set of ($\Mpbh, \lambda_E$) values satisfying the degeneracy condition, current CMB limits (\cref{fig:exclusion-plot}) require that PBH-powered AGN emit at significant super-Eddington ($\lambda_E \gtrsim 10$) luminosities. This constraint can be overcome if PBHs grow by $\gtrsim 10\times$ their initial mass by the time of observation. 

\item[{\color{red} $\blacksquare$}] In the PBH scenario, about 75\% of the observed galaxies with $\muv=-21$ at $z=11$ should host a PBH-powered AGN and typically reside in low mass halos, $M_h = 10^{8-9} \msun$ (\cref{fig:lum-fun}). These predictions can be thoroughly tested with available and forthcoming \textit{JWST} spectroscopic data.
\end{itemize}

\section*{Acknowledgments}

AF acknowledges support from the ERC Advanced Grant INTERSTELLAR H2020/740120.
Partial support (AF) from the Carl Friedrich von Siemens-Forschungspreis der Alexander von Humboldt-Stiftung Research Award is kindly acknowledged.
We gratefully acknowledge the computational resources of the Center for High Performance Computing (CHPC) at SNS.
We acknowledge usage of \code{Wolfram|Alpha}, the \code{Python} programming language \citep{python2,python3}, \code{Astropy} \citep{astropy}, \code{corner} \citep{corner},  \code{emcee} \citep{emcee}, \code{hmf} \citep{murray:2013}, \code{Matplotlib} \citep{matplotlib}, \code{NumPy} \citep{numpy}, and \code{SciPy} \citep{scipy}.

\bibliographystyle{aa_url}
\bibliography{bib/bibliography, bib/codes}

\appendix
\section{PBH contribution to the power spectrum with lognormal IMF}\label{app:lognormal-calculations}

We take a lognormal as the fiducial model for the PBH mass function
\be\label{eq:lognormal-pbh-mf-appendix}
    \deriv{n(>M)}{\log M} = \frac{n_0}{\sqrt{2\pi}\sigma}\exp\left(-\frac{\log^2 M / \Mpbh}{2\sigma^2}\right)\,,
\ee
where $n_0$ is the number density scale; $n_0$ can be connected to $f\ped{PBH}$ from
\begin{subequations}
\begin{align}
    f\ped{PBH} \rho_c \Omega\ped{dm} &= \int_0^{+\infty}M\deriv{n}{M}\dd M =\\
     & = \frac{1}{\ln 10}\int_0^{+\infty}\deriv{n}{\log M}\dd M = \nonumber\\
     &= n_0 \Mpbh e^{0.5\sigma^2\ln^2 10}\,, \nonumber
\end{align}
which implies
\be
    n_0 = \frac{f\ped{PBH} \rho_c \Omega\ped{dm}}{\Mpbh} e^{-0.5\sigma^2\ln^2 10}.
\ee
\end{subequations}
The corresponding fraction of DM mass in PBH of mass $M$ is:
\be
    f(M) = \frac{M}{\rho_c\Omega\ped{dm}} \deriv{n(>M)}{M} = \frac{1}{\rho_c\Omega\ped{dm}}\deriv{n(>M)}{\ln M}.
\ee

Inverting the relation for $k_{crit}$ in \cref{eq:power-spectrum-pbh} combined with \cref{eq:pbh-fraction},
it is possible to define 
\be
    M(k) \equiv 2 \pi^2 \Omega_{dm} \frac{\rho_{c}}{k^3}
\ee
as the mass associated with a given critical wavenumber.
The power spectrum contribution can then be computed as follows:
\begin{align}\label{eq:power-lognormal}
    &P\ped{PBH}(k) = &\\
    &\frac{D^2 n_0}{\sqrt{2\pi}\ln10\sigma\rho_c^2\Omega\ped{dm}^2} \int_0^{M(k)}\exp\left(-\frac{(\log M' / \Mpbh)^2}{2\sigma^2}\right)M' \dd M' \stackrel{x = \log M'}{=}&\nonumber\\
    &\frac{D^2 n_0}{\sqrt{2\pi}\sigma\rho_c^2\Omega\ped{dm}^2}\int_{-\infty}^{\log M(k)} 100^x \exp\left(-\frac{(x-\log \Mpbh)^2}{2\sigma^2}\right) \dd x =& \nonumber\\
    &\frac{D^2 f\ped{PBH}}{2\rho_c\Omega\ped{dm}}\Mpbh e^{1.5 \sigma^2\ln^2 10}\left(1-\erf\left(\frac{\log \Mpbh /M(k)+2\sigma^2\ln10}{\sqrt{2} \sigma}\right)\right).& \nonumber
\end{align}
Although it is not explicit from \cref{eq:power-lognormal}, in the limit $\sigma\to0$ we effectively recover the term in \cref{eq:power-spectrum-pbh}.
This expression is also convenient to avoid numerical problems with the derivation of the power spectrum around the scale of $k_{crit}$.

\section{Alternative parametrization of results}\label{app:second-corner}

\begin{figure}
    \centering
    \includegraphics[width=0.99\linewidth]{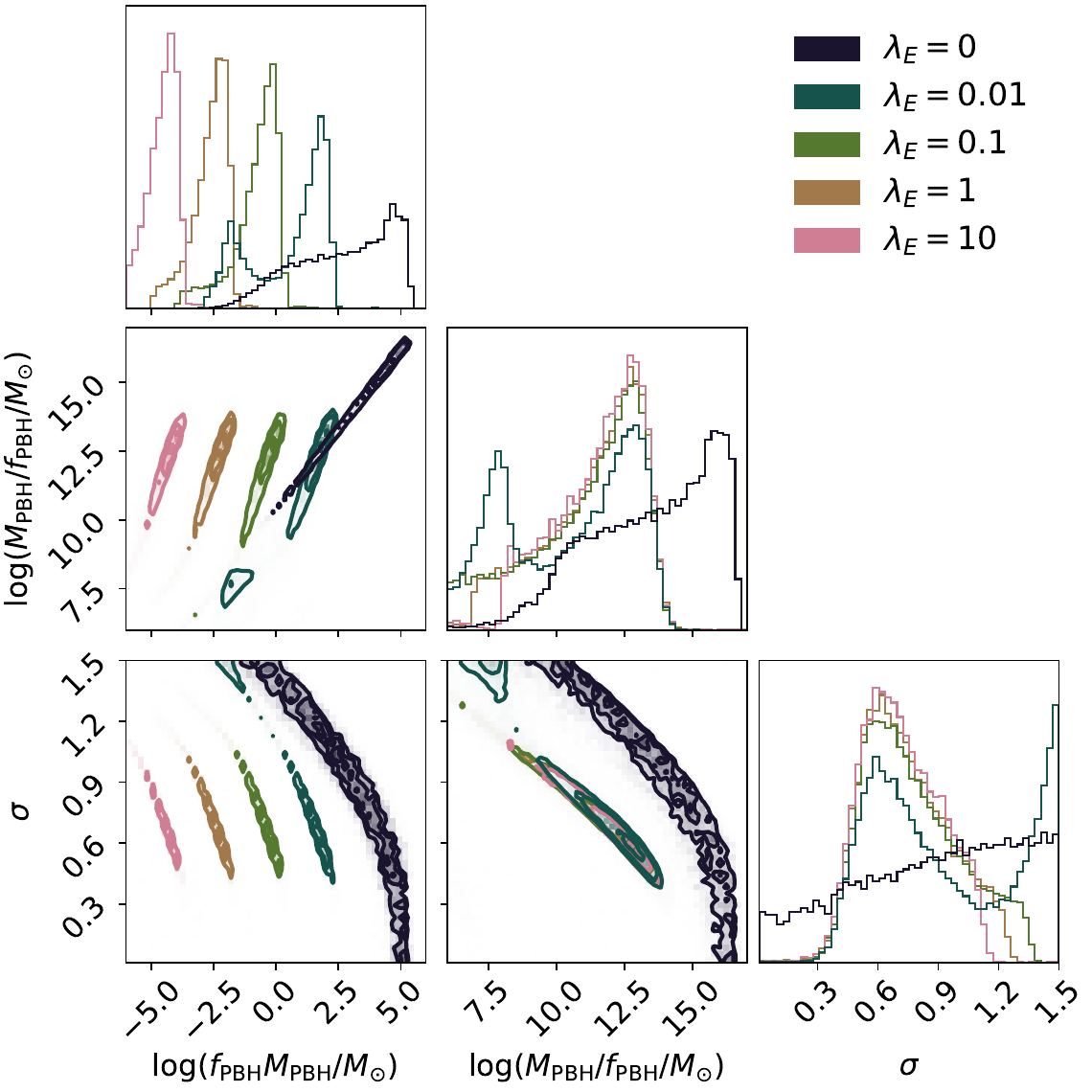}
    \caption{
        Alternative corner plot of the posterior for the PBH models (\cref{sec:method}).
        Parameters resulting from the MCMC procedure (see \cref{fig:corner-plot} and \cref{tab:emitting-pbh})
        have been recombined in the form of $\log \fpbh\Mpbh/\msun$ and $\log \Mpbh/\fpbh/\msun$.
        Each color corresponds to a value for $\lambda_E$, as reported in the legend.
        Tabulated results are reported in \cref{tab:alternative-results}.}
    \label{fig:second-corner}
\end{figure}

In the case of a monochromatic mass function for PBHs, the power spectrum enhancement is completely
degenerate in $\fpbh$ and $\mpbh$. For this reason, in \cref{fig:second-corner} and \cref{tab:alternative-results}
we report the results from \cref{sec:results} with an
alternative parametrization (product and ratio of $\fpbh$ and $\Mpbh$) that may be useful to compare with other works.

\begin{table}[t]
    \centering
    \begin{tabular}{c | c c c}
        \toprule
        $\lambda_E$ & $\log (\fpbh \Mpbh/\msun)$ & $\log (\Mpbh / \fpbh / \msun)$ & $\sigma$\\
        \midrule
        10   & $-4.46^{+0.47}_{-0.72}$ & $11.86^{+1.23}_{-2.17}$ & $0.70^{+0.23}_{-0.18}$ \\
        1    & $-2.50^{+0.50}_{-0.86}$ & $11.72^{+1.34}_{-2.60}$ & $0.71^{+0.27}_{-0.18}$ \\
        0.1  & $-0.56^{+0.53}_{-1.07}$ & $11.55^{+1.45}_{-3.10}$ & $0.73^{+0.31}_{-0.20}$ \\
        0.01 & $1.11^{+0.81}_{-2.74}$  & $10.83^{+2.10}_{-3.31}$ & $0.83^{+0.57}_{-0.27}$ \\
        0    & $2.94^{+1.78}_{-2.47}$  & $13.83^{+2.06}_{-2.92}$ & $0.93^{+0.39}_{-0.51}$ \\
        \bottomrule
    \end{tabular}
    \caption{
        Tabulated MCMC results for the alternative parameters whose corner plot is reported in \cref{fig:second-corner}.
        For each parameter, the fiducial value corresponds to the median, the uncertainties are built using the 16 and 84 percentiles.
        \label{tab:alternative-results}
    }
\end{table}

\label{LastPage}
\end{document}